\documentclass[RNAAS]{aastex63}

\submitjournal{RNAAS}

\shorttitle{WISE J135501.90-825838.9: A Young, Extremely Low-mass Substellar Binary}
\shortauthors{Theissen et al.}
\graphicspath{{./}{figures/}}

\begin{document}

\title{WISE J135501.90-825838.9 is a Nearby, Young, Extremely Low-mass Substellar Binary}

\correspondingauthor{Christopher A. Theissen}
\email{ctheissen@ucsd.edu}

\author[0000-0002-9807-5435]{Christopher A. Theissen}
\altaffiliation{NASA Sagan Fellow}
\affiliation{Center for Astrophysics and Space Science, University of California San Diego, La Jolla, CA 92093, USA}

\author[0000-0001-8170-7072]{Daniella C. Bardalez Gagliuffi}
\affiliation{American Museum of Natural History, 200 Central Park West, New York, NY 10024, USA}

\author[0000-0001-6251-0573]{Jacqueline K. Faherty}
\affiliation{American Museum of Natural History, 200 Central Park West, New York, NY 10024, USA}

\author[0000-0002-2592-9612]{Jonathan Gagn\'e}
\affiliation{
Institut de Recherche sur les Exoplan\`etes, Universit\'e de Montr\'eal, Pavillon Roger-Gaudry, PO Box 6128 Centre-Ville STN, Montr\'eal
QC H3C 3J7, Canada}

\author[0000-0002-6523-9536]{Adam Burgasser}
\affiliation{Center for Astrophysics and Space Science, University of California San Diego, La Jolla, CA 92093, USA}




\begin{abstract}
We present a parallax solution for WISE J135501.90-825838.9, a spectral binary with spectral types L7+T7.5 and candidate AB Doradus member. Using \textit{WISE} astrometry, we obtain a distance of $d = 16.7\pm5.3$~pc. This preliminary parallax solution provides further evidence that WISE J135501.90-825838.9 is a member of AB Doradus (130--200 Myr), and when combined with evolutionary models predicts masses of 11~$M_\mathrm{Jup}$ and 9~$M_\mathrm{Jup}$ for both components. 
\end{abstract}



\section{Introduction}\label{sec:intro}
Very low-mass ($M<0.1\,M_\odot$), young ($\lesssim 300$ Myr) binaries are important benchmarks for evolutionary models. Dynamical masses can help break the mass-temperature-age degeneracy due to the constant cooling of objects not massive enough to fuse hydrogen. The lowest-mass known binary is 2MASS J11193254-1137466AB, with a total estimated mass of $\sim 7.4\,M_\mathrm{Jup}$ and an estimated age of $\sim 10$ Myr \citep{best:2017:l4}. Systems such as these are extremely rare, with no estimates on their occurrence rate due to the difficulty in locating and identifying extremely faint, very low-mass binaries.

\citet{bardalez-gagliuffi:2018:101} identified WISE J135501.90-825838.9 (hereafter WISE J1355-8258) as a spectral binary, an object with a blended light near-infrared spectrum of a late-M/L primary and a T dwarf secondary \citep{burgasser:2010:1142, bardalez-gagliuffi:2014:143}. The estimated spectral types of the components are L6--7 for the primary, and T3--7.5 for the secondary. The largest source of uncertainty in the temperature/spectral type estimate is the age of the system, which is required to constrain the mass-temperature-age degeneracy. Using the Bayesian Analysis for Nearby Young AssociatioNs II \citep{malo:2013:88, gagne:2014:121}, \citet{bardalez-gagliuffi:2018:101} found that the $UVW$ kinematics and position of WISE J1355-8258 are consistent with the AB Doradus moving group \citep[130--200~Myr;][]{bell:2015:593} with a $>95\%$ probability. However, their age determination was based on a spectrophotometric distance estimate. This is problematic as the distance depends on the assumed luminosity of the object, which changes with age, and with the assumption of binarity. A trigonometric parallax distance, which is model-independent, would help inform the true age/luminosity of this system. This object does not have an entry in \textit{Gaia} Data Release 2 \citep[DR2;][]{gaia-collaboration:2018:a1}, likely due to its extreme optical faintness ($G > 21$ mag). Here we present a trigonometric parallax solution for WISE J1355-8258 using astrometry from the \textit{Wide-field Infrared Survey Explorer} \citep[\textit{WISE};][]{wright:2010:1868}.

\section{Methods and Discussion} \label{sec:methods}
\begin{figure}
\centering
\includegraphics[]{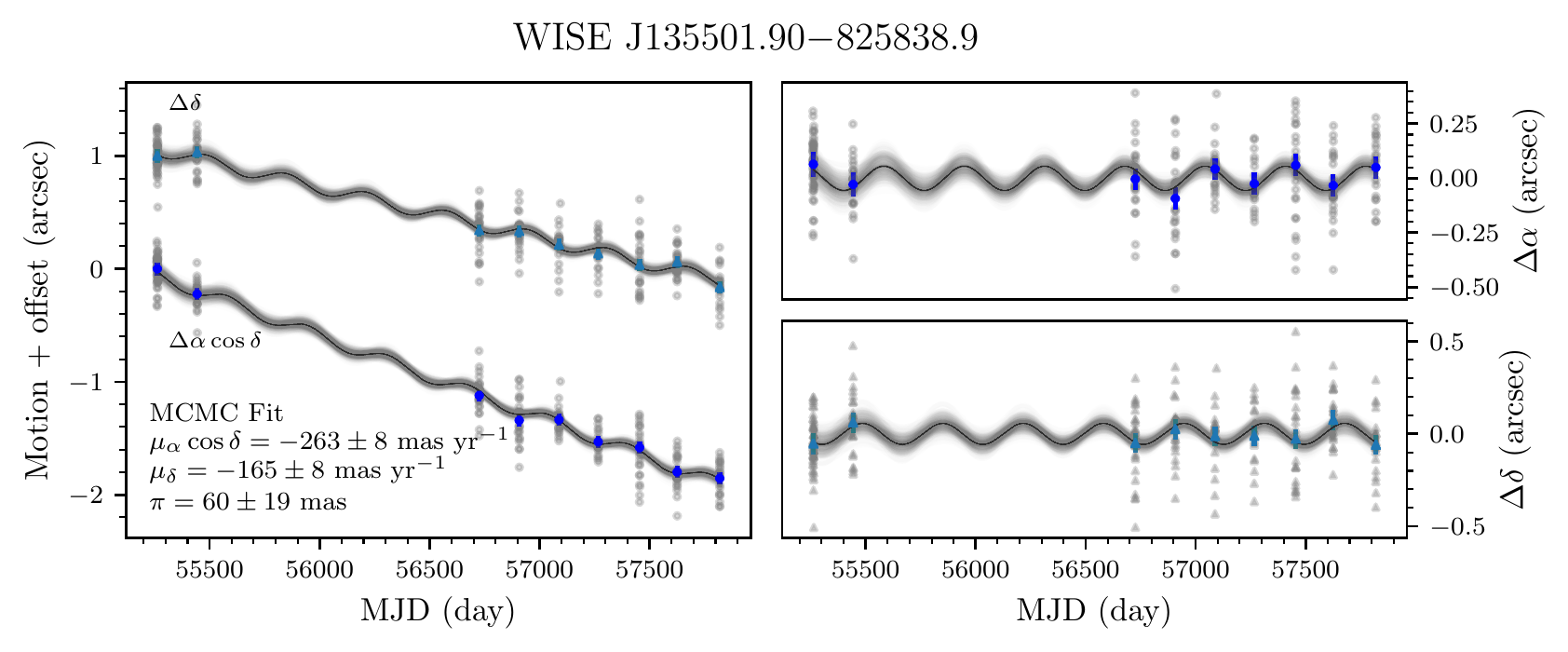}
\caption{\textit{WISE} parallax solution for WISE J1355-8258 (solid lines). \textit{Left:} The $\alpha$ and $\delta$ solutions are offset by 1\arcsec\ for visibility. Individual positions for each exposure are shown as translucent gray points. Blue points and cyan triangles display the uncertainty-weighted mean positions for each epoch. The light gray bands show 300 random realizations from the MCMC posterior distributions. \textit{Right:} the astrometric solution with the proper motions removed. The distance is constrained to $d = 16.7 \pm 5.3$ pc, which favors a young, extremely low-mass binary \citep[$M_\mathrm{tot} \lesssim 20~M_\mathrm{Jup}$;][]{bardalez-gagliuffi:2018:101}. 
}
\label{fig:parallax}
\end{figure}

We used the single-epoch astrometry from the Level 1b frames from the original \textit{WISE} mission and the reactivated NEOWISE mission \citep{mainzer:2011:53,mainzer:2014:30} to compute the parallax solution for WISE J1355-8258, as outlined in \citet{theissen:2018:173}\footnote{\url{https://github.com/ctheissen/WISE_Parallaxes}}. Figure~\ref{fig:parallax} shows the best-fit solution for the parallax and proper motion, constraining the parallax to $60\pm19$~mas, or $d=16.7 \pm 5.3$~pc. This solution is consistent with the with Case 3 from \citet{bardalez-gagliuffi:2018:101}, where the spectral types of each component are L7 and T7.5 with an estimated age of 130--200~Myr (from AB Doradus membership), a spectrophotometric distance of $d = 17\pm2$ pc, and estimated masses of $11\,M_\mathrm{Jup}$ and $9\,M_\mathrm{Jup}$, respectively. Using the updated parallax with the kinematics information from \citet{bardalez-gagliuffi:2018:101}, we used BANYAN $\Sigma$ \citep{gagne:2018:23} to reassess the probability that WISE J1355-8258 is a member of AB Doradus. We find using BANYAN $\Sigma$ that WISE J1355-8258 has a 95.6\% probablity of being a member of AB Doradus, a 3.4\% probability of being a member of $\beta$ Pic, and a 1\% probability of being a field object.

With a total mass of $\sim 20~M_\mathrm{Jup}$, WISE J1355-8258 is the second lowest mass binary known, with 2MASS~J11193254-1137466AB being the lowest-mass binary currently known \citep[$M_\mathrm{tot}\approx 7.4\,M_\mathrm{Jup}$;][]{best:2017:l4}. Our parallax solution also disfavors Case 1 ($M_\mathrm{tot} \approx 133\,M_\mathrm{Jup}$) and Case 2 ($M_\mathrm{tot} \approx 112\,M_\mathrm{Jup}$) from \citet{bardalez-gagliuffi:2018:101} at the 3-$\sigma$ and 2-$\sigma$ levels, respectively. A more precise parallax solution is required to definitively rule out higher-mass components. 

\acknowledgments

C.A.T. acknowledges support from NASA through the NASA Hubble Fellowship grant HST-HF2-51447.001-A awarded by the Space Telescope Science Institute, which is operated by the Association of Universities for Research in Astronomy, Inc., for NASA, under contract NAS5-26555.

This publication makes use of data products from the \textit{Wide-field Infrared Survey Explorer}, which is a joint project of the University of California, Los Angeles, and the Jet Propulsion Laboratory/California Institute of Technology, funded by the National Aeronautics and Space Administration.

This research made use of Astropy, a community-developed core Python package for Astronomy \citep{astropy-collaboration:2013:a33}. Plots in this publication were made using Matplotlib \citep{hunter:2007:90}. 

%

\vspace{5mm}
\facilities{IRSA, \textit{WISE}}


\software{Astropy \citep{astropy-collaboration:2013:a33},  
          Matplotlib \citep{hunter:2007:90},
          emcee \citep{foreman-mackey:2013:306},
          \textit{WISE} Parallaxes \citep{theissen:2018:173},
          }




\bibliography{MyLibrary.bib}
\bibliographystyle{aasjournal}



\end{document}